\documentclass[preprint,12pt]{elsarticle}

\AtBeginDocument{%
  \providecommand\BibTeX{{%
    \normalfont B\kern-0.5em{\scshape i\kern-0.25em b}\kern-0.8em\TeX}}}
\usepackage{soul}
\usepackage{amsfonts}
\usepackage{setspace}
\usepackage{array}
\usepackage{float}
\usepackage{amssymb}
\usepackage{amsmath}
\usepackage{graphicx}
\usepackage{url}
\usepackage{multicol}
\usepackage{stfloats}
\usepackage{caption}
\usepackage{subcaption}
\usepackage{placeins}
\usepackage{flushend}
\usepackage{hyperref}
\usepackage{booktabs}
\usepackage{changes}
\usepackage{eurosym}
\usepackage{rotating}
\usepackage{color}
\usepackage{xcolor}
\usepackage{makecell}
\usepackage{multirow}
\usepackage[]{xcolor}
\let\printorcid\relax 
\usepackage{graphicx}
\usepackage{algorithm}
\usepackage{algorithmic}
\graphicspath{ {Figures/} }

\newcommand{\bi}{\begin{itemize}}
	\newcommand{\ei}{\end{itemize}}
\newcommand {\beq}{\begin{equation}}
\newcommand {\eeq}{\end{equation}}
\newcommand {\be}{\begin{enumerate}}
	\newcommand {\ee}{\end{enumerate}}

\newcounter{RZNumberOfComments}
\stepcounter{RZNumberOfComments}

\begin{document}
	\let\WriteBookmarks\relax
	\def\floatpagepagefraction{1}
	\def\textpagefraction{.001}
	\title {Adversarial Botometer: Adversarial Analysis for Social Bot Detection}
        \author[ut,ua]{S. Najari}
        \ead{najari.shaghayegh@ut.ac.ir, najarigh@ualberta.ca}
        \author[ua]{D. Rafiei}
        \ead{drafiei@ualberta.ca}
        \author[ut,ipm]{M. Salehi\corref{cor}}
        \ead{Mostafa_salehi@ut.ac.ir}
        \author[els]{R. Farahbakhsh}
        \ead{reza.farahbakhsh@it-sudparis.eu}
        \cortext[cor]{Corresponding author}
        \address[ut]{Faculty of New Sciences and Technologies, University of Tehran, Tehran, Iran}
        \address[ua]{Computing Science Department, University of Alberta, Edmonton, Alberta}
        \address[ipm]{School of Computer Science, Institute for Research in Fundamental Science (IPM), P.o.Box 19395-5746, Tehran, Iran}
        \address[els]{Institut Polytechnique de Paris, Telecom SudParis, Evry, France}

        \begin{abstract}
		Social bots play a significant role in many online social networks (OSN) as they imitate human behavior.
This fact raises difficult questions about their capabilities and potential risks.
Given the recent advances in Generative AI (GenAI), social bots are capable of producing highly realistic and complex content that mimics human creativity.
As the malicious social bots emerge to deceive people with their unrealistic content, identifying them and distinguishing the content they produce has become an actual challenge for numerous social platforms.
Several approaches to this problem have already been proposed in the literature, but the proposed solutions have not been widely evaluated. To address this issue, we evaluate the behavior of a text-based bot detector in a competitive environment where some scenarios are proposed:
\textit{First}, the tug-of-war between a bot and a bot detector is examined. It is interesting to analyze which party is more likely to prevail and which circumstances influence these expectations. In this regard, we model the problem as a synthetic adversarial game in which a conversational bot and a bot detector are engaged in strategic online interactions. \textit{Second}, the bot detection model is evaluated under attack examples generated by a social bot; to this end, we poison the dataset with attack examples and evaluate the model performance under this condition. \textit{Finally}, to investigate the impact of the dataset, a cross-domain analysis is performed. Through our comprehensive evaluation of different categories of social bots using two benchmark datasets, we were able to demonstrate some achivement that could be utilized in future works.
	\end{abstract}
	
	\maketitle

	\section{Introduction} \label{sec:Introduction}
Along with the development of Artificial Intelligence (AI) and since it enters to bots like Joseph Weizenbaum's ELIZA for emulating a Rogerian psychotherapist, many things have changed

%The history of chatbots goes back to as early as 1960s when Weizenbaum developed one of the first rule-based bots for emulating a Rogerian psychotherapist \cite{weizenbaum1966eliza}.
%There has been many changes since then, with major progress made in AI and machine learning and the reduced cost of computing resources and data acquisition.
The utilization of these automated algorithms in social networks, commonly known as social bots may initially have some positive effects; however, as time passes, their activities can become increasingly destructive.

In 2017, the average presence of bots on active Twitter accounts was estimated to be around 15\% \cite{varol2017online}, while on Facebook, it was approximately 11\% in 2019 \cite{zago2019screening}. These numbers indicate a considerable share of automated accounts on both platforms. Moreover, the presence of bots tends to increase significantly when there are strong political or economic interests involved. A study conducted in 2019 revealed that 71\% of Twitter users discussing trending US stocks were likely to be bots \cite{cresci2019cashtag}. Similar results were found regarding the presence of bots in online cryptocurrency discussions \cite{nizzoli2020charting} and their involvement in spreading “infodemics" during the COVID-19 pandemic \cite{gallotti2020assessing}.

In recent years, the rapid development of new models in GenAI leads to the emergence of powerful transformer-based bots such as Generative Pre-trained Transformer (GPT) \cite{vaswani2017attention}. %\reza{better to cite GPT4 and chatGPT which are more recent, and also add few words  for the new buzz around chatGPT}. 
These advancements have enabled social bots to engage in more complex interactions and penetrate popular discussions, such as participating in entertaining conversations, leaving comments on posts, and responding to questions \cite{ferrara2016rise}.
%%%%%% Previous studies

In response to detrimental activities of social bots, extensive research has been devoted to identification and mitigation of social bots.
%With the rapid development of new models in GenAI, the social bots are able to produce highly realistic and complex content that mimics human creativity. Despite this rapid development of social bots, bot detection techniques continue to encounter significant challenges that considerably restrict their practical applications.
%
%\subsection{Research GAP}
However, a major challenge with bot detection is the poor performance of the models under more complex circumstances, especially when a social bot employs deceptive tactics. This is mainly because a social bot detector is considered a fixed module without any progress, and thus the data on which the model is trained on is also considered fixed \cite{DENICOLA2021102685}. This is problematic when the data that the model is trained on fails to capture unforeseen or future patterns. Even a small irregularity in the training data can cause the model's performance to drop significantly.

Evaluating the behavior of bot detection models in the presence of attack examples generated by human-like bots is an under-researched area.
This study aims to address this gap by exploring a novel approach wherein a bot engages in an adversarial game with a bot detection model.
%In this work, we consider a bot interacting with bot detection through an adversarial game. 
In this game, where a bot automatically generates perturbations, the performance of bot detection model can be evaluated using a process called adversarial training \cite{goodfellow2014explaining}.
%
%\subsection{Contributions}
In particular, we formulate adversarial scenarios in which a bot simulates human behavior to generate attack examples. In contrast, a bot detection model is then tasked with distinguishing between real and fake examples. With this approach, we can effectively evaluate the behavior of bot detection models to detect and defend against attacks by fraudulent bots.

Our evaluation, conducted on two benchmark datasets: Midterm-2018 and Cresci-2017 that includes three different categories of social bots and one category of human users. Through analyzing the results, we have reached some achievements that could utilize for the future works.

The main contributions of this study can be summarized as follows:
\begin{itemize}
\item We model a social bot as an interactive and automated conversational model.

\item 
To evaluate the bot detector's behavior in a dynamic condition, we design an adversarial game between bot detector and the bot that is producing some adversarial attacks.
%As the goal of a social bot is to remain hidden from the detectors, we develop a live adversarial game between the designed social bot and a text-based bot detector (contextual-LSTM \cite{kudugunta2018deep}), to do a competition with each other.

\item To thoroughly evaluate the performance of the bot detector, we ran 3 different scenarios and presented our achievements, which we can use for future work.
\end{itemize}

%For our evaluation, we use a benchmark dataset that includes three different categories of social bots and one category of human users. By evaluating the accuracy of our model in different situations, we show that text-based social bot detectors are not reliable enough in the scenarios we study. This highlights the need for further study on this topic as well as to develop bot detection techniques that address the challenges posed by the different behaviors of bots.

The rest of the paper is organized as follows. In Section \ref{sec:RelatedWorks}, we review the literature related to our work. Section \ref{sec:ProposedMethod} presents our proposed model and scenarios and Section \ref{sec:Experiments} presents our results and analysis. Finally, we conclude the work with a discussion of some future research directions in Section \ref{sec:Conclusions}.
	\section{Related Works} \label{sec:RelatedWorks}
To evaluate the performance of the bot detection module when faced with a complex social bot that mimics human behavior, it is important to have an overview of generative bots and the bot detectors. Therefore, we will discuss them in the following sections.
\subsection{Bot Generation}
As automated computer programs, bots have existed in various forms since the early days of computers. These forms range from those controlled by humans, such as spam generators \cite{ wang2010detecting}, to fully automated algorithms like chatbots.
%%%%%%%%%%%%%%%%%%%%Dialog models
Advances in natural language processing, particularly the use of simple neural networks and transfer learning-based models \cite{weiss2016survey}, have enabled bots to engage with real users and generate text that closely resembles human text. This has led to a major challenge in distinguishing between content generated by bots and content created by humans \cite{alarifi2016twitter}.

One type of automated generative models that is designed to behave like a human is a dialogue model \cite{xu2021topic} .
This is a model that is able to capture the structure and meaning of conversations between two entities, typically a user and an interactive computer system.
Dialogue systems can be categorized into two main classes \cite{chen2017survey}: task-oriented and non-task-oriented (also referred to as open-ended conversational agents). Task-oriented dialogue systems are developed to accomplish specific tasks, whereas non-task-oriented systems are more versatile and capable of engaging in broader and more general conversations.
A widely adopted approach for constructing dialogue models is the utilization of a sequence-to-sequence model \cite{li2015diversity}. This model comprises an encoder component responsible for mapping the input sequence to an intermediate vector, and a decoder component that generates a response utilizing the hidden state of the encoder and the intermediate information obtained.

%%%%%%%%%%%%%%%%%%%%%%%%%%%%%%%%GAN
In the recent decade, Generative Adversarial Network (GAN) as a generative model designed for sequential data such as image generation \cite{goodfellow2014generative}, some later they were adapted to process discrete and textual data. SeqGAN is an example of a GAN, which has been specifically adapted for text generation \cite{yu2017seqgan}.
Here, both the generator and the discriminator work together to improve the quality of the text produced by the generator. The generator makes decisions at each time step to maximize expected rewards, which are determined by the discriminator. \cite{yu2017seqgan}. 

As the goal of this paper is to analyze the behavior of bot and bot detection in a tug-of-war, we utilized the dialogue models beside SeqGAN to design some competition scenarios based on.
%%%%%%%%%%%%%%%%%%%%%%%%%%%%% Adversarial Attack

\subsection{Bot Detection}
In general, contributions to analyze the bot detectors behavior could to fall into two main categories: feature-based and model-based approaches.

In Feature-based category, social bots are identified by some simple and well-known machine learning models, such as Random Forest and Support Vector Machine, by analysing a set of behavioural features extracted from the input dataset \cite{heidari2021empirical, aljabri2023machine, orabi2020detection}, directly. Thus, the performance of these models are more dependent on the features selected. 

Features can be divided into two main classes: content-based and user-based. Content-based features focus on the content of the behaviour-based post, including the text of a tweet or linguistic annotations such as part-of-speech tags \cite{jr2018detection}.
User-based features, on the other hand, are based on a user's profile, behavior \cite{velayutham2017bot}, and  connections to other users in the network \cite{dorri2018socialbothunter}.

In another category, models are improving to extract more informative features from the original contexts.
Recently, some models such as deep learning-based techniques have emerged that attempt to improve social bot detectors by automatically extracting features and feeding them into a deep neural network that enables the extraction of more informative feature vectors. By using multiple layers of interconnected neurons, deep learning models can capture intricate patterns and representations in the data, leading to more effective feature extraction \cite{hayawi2023social}. 
 
Long Short-Term Memory (LSTM) and Convolutional Neural Networks (CNN) are popular neural network architectures that have achieved successful results in tasks such as bot detection \cite{kudugunta2018deep, beskow2019its, arin2023deep}. In this area, a novel deep neural network model named RGA (ResNet, GRU, and Attention) is developed to address the task of detecting social bots. The RGA model combines the power of a residual network (ResNet), a bidirectional gated recurrent unit (BiGRU), and an attention mechanism \cite{wu2021novel}. These methods often represent the raw text as a vector and use pre-trained linguistic models such as Global Vectors for Word Representation (GloVE \cite{pennington2014glove}) to encode initial features.

 \subsection{Existing Research GAP and Objective}
With the advance of bots through recent developments in genAI, there is a pressing need for powerful social bot detection methods in social networks to understand better the influence and impact of these bots and develop effective countermeasures \cite{ferrara2023social}.

As previous studies have shown, there is a lack of effective studies to evaluate social bot detection methods when confronted with intelligent bots that attempt to deceive the detectors. This was a strong incentive for us to take a new perspective and try to address this problem in more detail.

Most of the previous studies use the GAN framework to augment the lower class data such as bot \cite{najari2022ganbot}, fake reviews \cite{aghakhani2018detecting}, or spam \cite{shehnepoor2020gangster} to make the detectors stronger.
Here, we would like to use GAN not as an auxiliary module but as an adversary module to design a game between bot and bot detector to analyze their behavior and strength under dynamic conditions.

Specifically, we use GAN to extend the bot class not by bot-like samples, but to include examples generated by human or human-like models: Human-like examples are generated by the adversarial bot (pre-trained on the class bot and fine-tuned on the class human), which knows the bot detection results as feedback. This extension will allow the evaluation of bot detection models in terms of their ability to distinguish between bots and humans in more complex states.
In this study, we focus on evaluating the effectiveness of bot detection models in the presence of attack examples produced by generative model. 
	
	\section{Proposed  Method: Adversarial Botometer}\label{sec:ProposedMethod}
In this part, we have review the proposed scenarios. 
%\reza{well better not to call it scenario in the title, if we have a name for the new model we call it here, if not we say proposed approach based on Adversarial Game}
\subsection{Scenario 1: Adversarial Game between bot and bot bot detection}
\begin{figure}
	\includegraphics[width=1\linewidth]{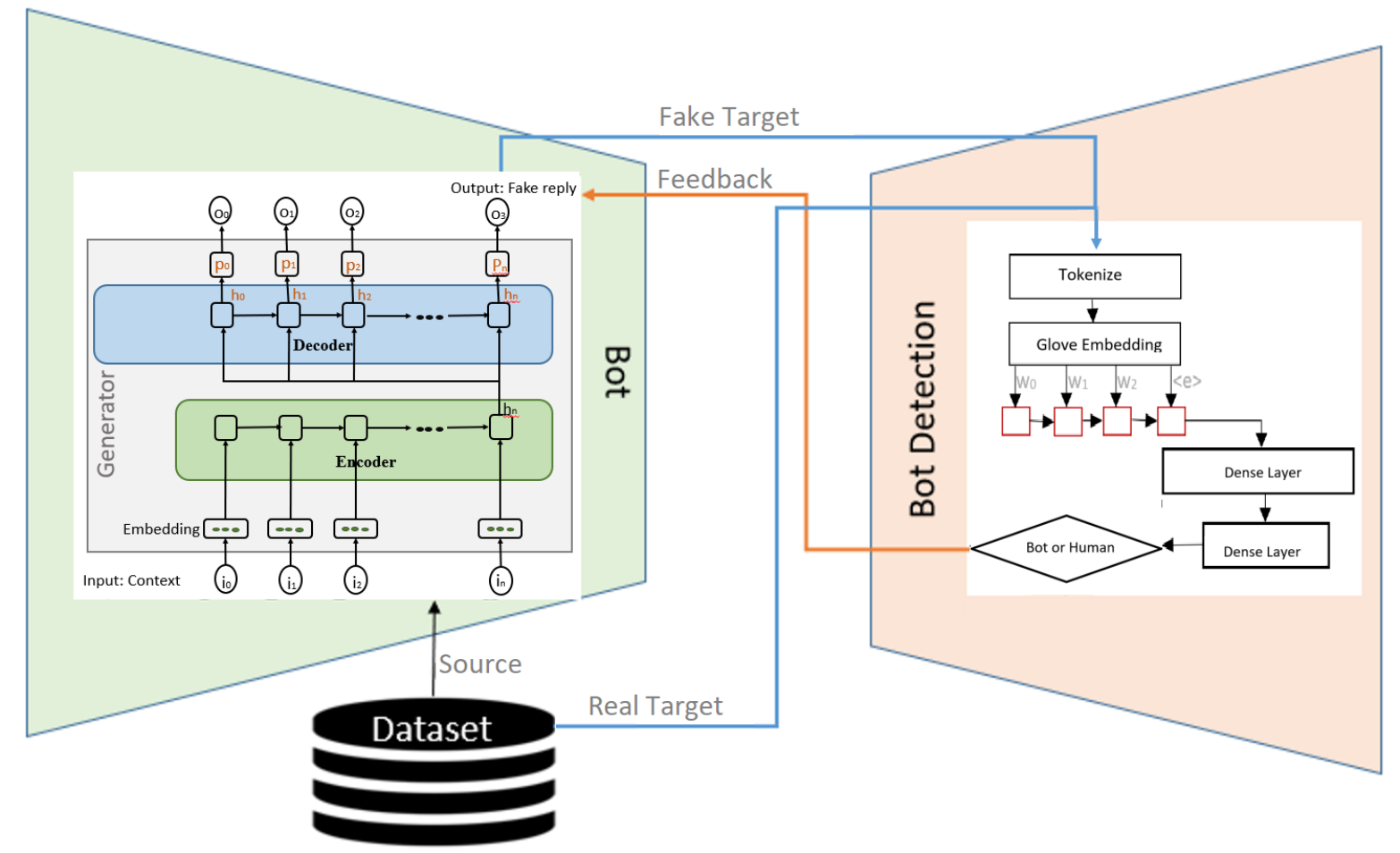}
	\caption{Proposed Adversarial Game}
	\label{fig: ADV-SBD}
\end{figure}

Figure \ref{fig: ADV-SBD} presents an overview of our proposed framework, which entails the design of an adversarial game between a bot and bot detection model. In this setup implemented by GAN, the bot is responsible for generating samples that resemble human behavior. On the other hand, the bot detection model tries to distinguish between the patterns generated by the bot and those generated by humans. Through this adversarial interaction, the generative model seeks to improve its ability to generate samples that are indistinguishable from those generated by humans, while the discriminator aims to accurately identify bot-generated samples.

%, in our case, using a contextual-LSTM \cite{kudugunta2018deep} (a type of recurrent neural network) 
% adversarial game between the bot and bot detection models.
To design this synthetic game, it is first necessary to specify a generative model that plays the role of the bot, and a discriminator model that plays the role of the bot detector. As shown in the previous section, there are a wide range of models to do this selection. Here, we select the Seq2Seq model as the generative model and the Contextual-LSTM as the discriminator model for several reasons:
\begin{itemize}
    \item {\textbf{Bot Generation: Seq2Seq Model}} 
    
    With the growth of bots through recent developments in GenAI, the recent social bots are not rigid generative models; rather, they are able to interact with other users, mimic their behaviour. To address this, we chose the seq2seq model as the basis for dialogue models.
    \item {\textbf{Bot Detection: Contextual-LSTM}}
    
    Since in our problem a bot tries to generate text like a human, the generated content is text. As the text is sequential data and LSTM is known to be the best example of DL-based models for sequential data, we chose Contextual-LSTM as the basis for text-based bot detectors.
\end{itemize}

The training phase for the bot and bot detection models consists of two main steps: Pre-training, and Adversarial-training.

\subsubsection{Pre-training Phase}
In the context of dialog settings, our main focus is on generating answers to questions or comments based on observed posts. To achieve this, we use a sequence-to-sequence model consisting of an encoder and a decoder.

The encoder component captures an encoded representation of the input question or comment to help the model to understand the input and extract relevant features.
The decoder component takes the encoded representation from the encoder and generates a response, in return. 

Formally speaking, given a corpus of message pairs $(S, R)$ where the source message $S$, consists of tokens ${s_1, ..., s_n}$ and the response (target) message $R$, consists of tokens ${r_1, ..., r_m}$, the bot is trained to maximize the total log likelihood of observed target messages, given their respective source messages:
\begin{equation}
	\sum_{(S,R)} \log P(r_1,...,r_m|s_1,...,s_n)
\end{equation}
Here, the goal of decoder $D$ is to find an approximated distribution of trainset conditioned on generator parameter $\theta$, and initial source $S$. Then, the decoder produces each token of samples based on preceding ones as shown in eq.\ref{eq:next-token}.
\begin{center}
	\vspace{-0.7cm}
	\begin{equation}
		D_{\theta}({R_{1:T}|\theta, S})=\displaystyle\prod_{t:1,2,...,T} D_{\theta}(r_t|r_{1:t-1}, \theta, S)
		\label{eq:next-token}
	\end{equation}
\end{center} 

As illustrated in Figure \ref{fig: ADV-SBD}, the bot detector takes two types of inputs: fake targets generated by the generative model, or real ones selected from the target dataset.
Given a corpus of pair message $(S,R)$, score $y=1$ if $R$ was sampled from the training data and $y=0$ otherwise, this model is trained to maximize the probability of correct labeling as follows:
\begin{equation}
	\sum_{(y,S,R)} \log P(y|s_1,...,s_m,r_1,...,r_n)
 \label{eq:pretrain}
\end{equation}

\subsubsection{Adversarial Training}
During the adversarial training phase, the goal is to generate samples that have higher realism, resulting in higher rewards. Here, a mismatch arises between the generative bot that produces sentences token by token and the discriminator model that evaluates the entire generated sequence.

To solve this problem, several approaches have been proposed previously. One effective method is to use a Monte Carlo Search Tree (MCST). MCST uses a tree-based search algorithm that explores the potential upcoming sequences \cite{yu2017seqgan}. The signaling feedbacks obtained from the discriminator can then be propagated backwards to the generator. The minimax objective function of this adversarial game could be as follows:
\begin{center}
	\begin{equation}
		\begin{split}
			min_B\_max_{BD} {( E_{(s,r)\sim P_{(S, R)}}\log{BD(r)}+\log{(1-BD(B(r|s)))} )}\\
			\label{eq:adv}
		\end{split}
	\end{equation}
\end{center}

Here, $B$ and $BD$ demonstrate the bot and bot detector models, respectively. $B$ generates samples satisfying $P_B(r|s)$; where, the generator $B$ as the bot generate sample $r$ selected from the response set $R$ and given $s$ selected from the source set $S$.
Practically, we use a Long Short Term Memory (LSTM) to generate words, where each recurrent unit has embedding size 25, hidden dimension and a batch size of 64.

In this game, the bot detector ($BD$) provide a feedback to the bot ($B$). This feedback serves as an indicator of the bot's success or failure in fooling the bot detector. Based on this information, the bot regulates its generation strategy and produces more patterns, which are then sent to the bot detector for evaluation. This iterative process continues in a loop, forming a dynamic game between the bot and the bot detection model.
% C, R - S,T- I,O
%\subsection{Training Process} 
Algorithm \ref{alg:alg1} provides more details about the process of this game.

%#################################3
As shown in  Algorithm \ref{alg:alg1}, the proposed framework first normalizes and pre-processes inputs \cite{kudugunta2018deep} in order to prepare tweets for input to the LSTM network. This pre-processing phase involves removing punctuation, tokenzing the tweets using the methods from Global Vectors for Word Representation (GloVE) \cite{pennington2014glove}%, and replacing certain elements including hashtags, URLs, numbers and user mentions with special tags such as $<hashtag>$, $<url>$ ,$<number>$ , and $<user>$, respectively. Other pre-processing steps, such as replacing emojis with the corresponding tags (\eg $<hear>$, $<smile>$, $<lol mode>$, $<neutral mode>$ and $<angry mode>$) are also performed; similar pre-processings are done in the related work\cite{kudugunta2018deep}.

We evaluate this framework through both live and offline adversarial game, and the results are reported in the next section.

\begin{algorithm}
	%\color{blue}
	%	\scriptsize
	\caption{Adversarial game of bot and bot detection models.}
	\begin{algorithmic}
		\REQUIRE Bot detector $BD$, Bot $B$, Dataset $X$ containing pairs of ($s$, $t$), as a symbol of (Source, Target)	
		\STATE Initialize $B$, $BD$ parameters with the random weights
		\STATE Pre-process and tokenize pairs ($s$, $t$) available in $X$
		\STATE Pre-train $B$ and $BD$
		\FOR {Training Iterations} 
		
		\FOR {$BD$-training-steps}
		\STATE Sample $t$ from the dataset $X$		
		\STATE Sample $\hat{t} \sim B(s)$\\
		\STATE Update $BD$ using ($t$) as positive sample and ($\hat{t}$) as negative sample using Eq. \ref{eq:pretrain}\\
		\ENDFOR
		\FOR {$B$-training-steps} 
		\STATE Sample ($s$,$t$) from the dataset $X$	
		\STATE Sample $\hat{t} \sim B(s)$\\
		\STATE Update $BD$ using ($t$) as positive sample and ($\hat{t}$) as negative sample and use output of $BD$ as reward $BD(B(s))$ in Eq. \ref{eq:adv}
		\STATE Update $B$ using defined objective function in Eq. \ref{eq:adv}
		\ENDFOR			
		\ENDFOR
	\end{algorithmic}\label{alg:alg1}
\end{algorithm}
\subsection{Scenario2: Data Poisoning} % Offline Adversarial Attacks:
To assess the effectiveness of our bot detection model on a more complicated dataset that include examples of attacks, we adopt two distinct approaches. Firstly, we can select attack examples directly from the dataset itself. Alternatively, we can generate attack examples by simulating bot behavior using the GAN framework.

In the first approach, we examine the existing dataset and specifically identify instances that illustrate attack behaviors. These examples may involve various forms of malicious activities, such as spamming, misinformation propagation, or coordinated manipulation. By incorporating these attack examples into our evaluation, we can evaluate the robustness and accuracy of our bot detection model in detecting and classifying these malicious behaviors.

The second approach is to generate attack patterns by simulating the behavior of the bot. Using the GAN technique and methods described earlier, we can simulate the actions and patterns of bots involved in attacks. By generating synthetic attack patterns, we can create a controlled environment to comprehensively evaluate the performance of our bot detection model in identifying and distinguishing between normal user behavior and malicious bot behavior.

Both approaches provide valuable insights into the effectiveness of our bot detection model in processing complex datasets of attack patterns. By combining real-world attacks with simulated attack scenarios, we can thoroughly evaluate the model's capabilities and improve its ability to detect and mitigate various forms of bot-driven attacks.

\subsection{Scenario3: Domain and Model Explanation}
To comprehensively evaluate the textual distinctions among social bots, we carried experiments on 14 NLP features extracted from each class of social bot datasets. These features included mention count, hashtag count, stopwords count, word count, unique word count, quoted word count, character count, sentence count, capital character count, capital word count, unique-to-total word ratio, average sentence length, average word length, and stopword-to-total word ratio.

In order to assess the significance of these features in prediction, we employed SHAP (SHAPley Additive Explanations) values to reverse-engineer the output of the predictive algorithm. This involved training the bot detection model on the aforementioned features from the three distinct classes of social bots. Subsequently, we evaluated the model's performance by testing it on the corresponding test set.

By utilizing SHAP values, we gain insights into the importance and contributions of each feature towards the model's predictions. This analysis enables us to understand the relative influence of different textual characteristics on the bot detection process. Through this approach, we can identify the key indicators and linguistic patterns that distinguish the various classes of social bots, further improving the effectiveness of our detection model.

	\section{ Experimental Results}
\label{sec:Experiments}
%To evaluate the effectiveness of our proposed methods, we did extensive experiments under various conditions. Our evaluation is performed on a dialogue dataset that includes interactions between humans and bots.
\subsection{Datasets}  \label{dataset}
In this study, we use a dataset of Twitter platform that serves as a reference point for spambot detection \cite{cresci2017paradigm}. This dataset includes two categories of accounts: genuine accounts and social bots. The genuine accounts managed by humans and social bots handeling by bots are divided into three subgroups: political bots (referred to as social spambots 1), financial bots (referred to as social spambots 2), and commercial bots (referred to as social spambots 3). To identify the political bots, we analyzed their activity of retweeting posts related to an Italian political candidate. The financial bots were identified as those responsible for spamming paid mobile apps. Finally, the commercial bots were identified as spammers promoting products on Amazon.com. We named these subgroups according to their main functions.

Here, we have extracted the conversations based on done replies on tweets by using the in\_reply\_to\_status\_id parameter available in the dataset.
\begin{table}
	\begin{center}
		\caption{Dataset - Cresci 2017 \cite{cresci2017paradigm} and Midtrerm 2018 \cite{yang2020scalable} .}
		%\reza{cite Cresci in the caption pls}
		\label{tb:dataset}
		\resizebox{\columnwidth}{!}{%
			\begin{tabular}{c|c|c|c}
				%	\textbf{} & \textbf{Tweet-Retweet} & \textbf{Tweet- Reply} \\ % <-- added & and content for each column
				%	$\alpha$ & $\beta$ & $\gamma$ & $\delta$ \\ % <--
				\textbf{Dataset} &\textbf{Label} & \textbf{\# Tweets} & \textbf{\#  Conversations}\\
				\hline
				\multirow{ 5}{*}{Cresci2017}&Human & 39,264 (38,516 , 748) & 16,967 (16593, 374)\\
				&Bot1-Political & 3810 (1,054,  2,756)& 1778 (400, 1378)\\
				&Bot2-Financial & 932 (932, 0)& 434 (434, 0) \\
				&Bot3-Commercial & 430 (176, 254) & 200 (73, 127) \\
				&Bot(Bot1+Bot2+Bot3) & 5172 (2,162 , 3,010) & 2,412 (907, 1,505 )\\
				\multirow{ 2}{*}{midterm-2018}&Human&8,092 & \\
				& Bot&42,446 & \\
			\end{tabular}
		}
	\end{center}
\end{table}
The statistical information of the Cresci's dataset is summarized in Table \ref{tb:dataset}. To prepare the dataset, we extracted Tweet-Retweet and Tweet-Reply relationships between users based on human-human and bot-bot interactions available in the dataset.
Also, we used the midterm dataset that is filtered based on political tweets collected during the 2018 U.S. midterm elections \cite{yang2020scalable}.
%\subsection{scenario1}
\begin{figure}[H]
\begin{center}
  \includegraphics[width=0.7\linewidth]{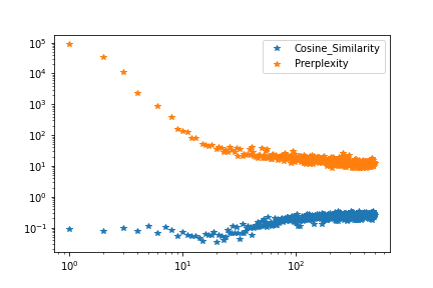}
  \caption{ Training procedure of bot in the context of GAN (Perplexity vs. Cosine similarity of  generated samples to the real ones}
  % \reza{pls add caption for x y axes}
  \label{fig: cs_pp}
\end{center}
\end{figure}

\subsection{scenario 1}
To gain insight into the behavior of our models and identify areas where performance degradation occurs and finally propose solutions, we carried adversarial training.

We used dialogue models to generate responses as a social bot, considering the context of ongoing conversations. On the other hand, we used a Contextual LSTM bot detection model to distinguish between machine-generated (social bots) and human-generated examples.

% In the adversarial training phase, the pre-trained bot tries to generate some samples similar to the human ones. Then, the generated samples besides some real samples generated by humans go forward to the bot detection model to distinguish if the input belongs to the actual class or generated one. 

In our adversarial scenario, our first goal was to evaluate the generative model's ability to produce samples that closely resembled real samples. We focused on evaluating how well the generated samples reproduce the features and patterns observed in the genuine data.

By comparing the generated samples to the real ones, we were able to determine the truth and realism of the generative model results based on GAN framework. This evaluation was important in determining the model's ability to reproduce the nuances and subtleties of the genuine data. It served as a fundamental benchmark for measuring the performance and accuracy of the generative model and provided valuable understanding into its capabilities. Figure \ref{fig: cs_pp} shows the result of this evaluation. The x-axis represents the number of training iterations for the generator model, while the y-axis represents the perplexity of the generated samples and the cosine similarity between the generated and actual samples. As can be seen in the figure, the cosine similarity between the generated samples and the real ones increases with the training of the bot model and the complexity of the bot decreases. This is to be expected as the generative model gets better at generating samples that can fool the discriminator model.

\begin{figure}
        \begin{subfigure}[ht]{0.45\linewidth}
            \includegraphics[width=5cm, height=4cm]{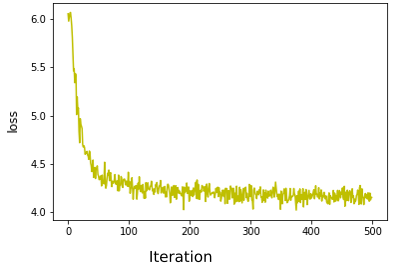}
            \caption{Bot detection Loss}
        \end{subfigure}
        \begin{subfigure}[ht]{0.45\linewidth}
            \includegraphics[width=5cm, height=4cm]{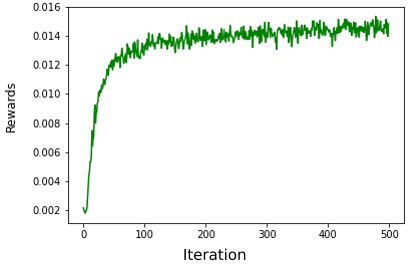}
            \caption{Rewards received from the bot detector to the bot}
        \end{subfigure}
        \caption{Training procedure of bot detection in the context of GAN.}
      %  \reza{add x y axes}
        \label{fig:dis_loss_reward}
\end{figure}
\begin{figure}[H]
\begin{center}
  \includegraphics[width=0.7\linewidth]{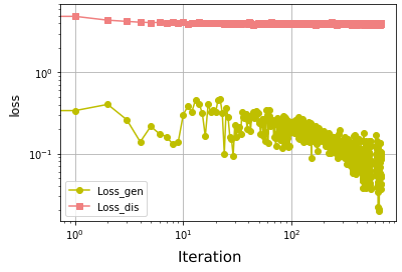}
  \caption{Training speed of bot and bot detection in the context of GAN}
  %\reza{can we have label for x-axes y-axes?}
  \label{fig:ADV-SBD}
 \end{center}
\end{figure}
Our next goal is to quantify the fluctuations of the bot detector druring the adversarial training process. Figure \ref{fig:dis_loss_reward} shows the loss and the rewards that the bot detector sends to the bot throughout the training process.

By analyzing the loss and reward curves, we represent the performance of  bot detector model and its ability to discriminate human and bot-generated samples.

The observed decreasing loss of the bot detection model in the context of GAN demonstrate its improvement during training. As the model optimizes its parameters, it becomes more skilled at discriminating between samples generated by social bots and those generated by humans. This indicates a positive trend in the model's ability to accurately detect machine-generated samples.

According to the GAN concept, the increasing signal rewards received to the social bot suggests that the bot detection model is improving at identifying and classifying bot-generated samples. So, the improvement in the bot detection is a direct consequence of the training process in the GAN framework.

Here, we modeled a GAN-based game between bot and bot detection models. As shown, each of bot and bot detection models (traind on the bot dataset) are working right like the generative and discriminator models in the context of GAN.

Now, we evaluate the training speed of the bot and bot detection models based on their training process in the designed competition. 
Earlier in the evaluation of bot detector's performance \ref{fig:dis_loss_reward} it can be observed that although the reward increases and the loss decreases, the plots in the later iterations exhibit smoother slopes. This trend suggests that the bot detection model encounters difficulty in extracting substantial information from the generated samples during these iterations. Moreover, the training loss of two models are compared in Figure \ref{fig:ADV-SBD}. Here, it is obvious that the training speed of the bot outperforms the performance of the bot detector model. In fact, the comparison of the loss values confirms the faster training speed of the generative model, indicating that it can outperform the bot detection model in terms of training efficiency. 

\textbf {Achievement 1:} In the GAN framework, a weak discriminator can be the reason of the other problems such as collapse mode, which can be solved by replacing autoencoders \cite{pinaya2020autoencoders}, Energy-based GANs \cite{zhao2016energy} or a strong cost function \cite{che2020your}. 

Since the behavior of the bot and the bot detector on the human-bot data is the same as the behavior of the generative and the discriminator models in the GAN context, we can use the solutions of the powerful discriminators proposed in the GAN framework for the bot detection method, too.
\begin{figure*}
\begin{subfigure}{0.3\textwidth}
    \includegraphics[width=\textwidth]{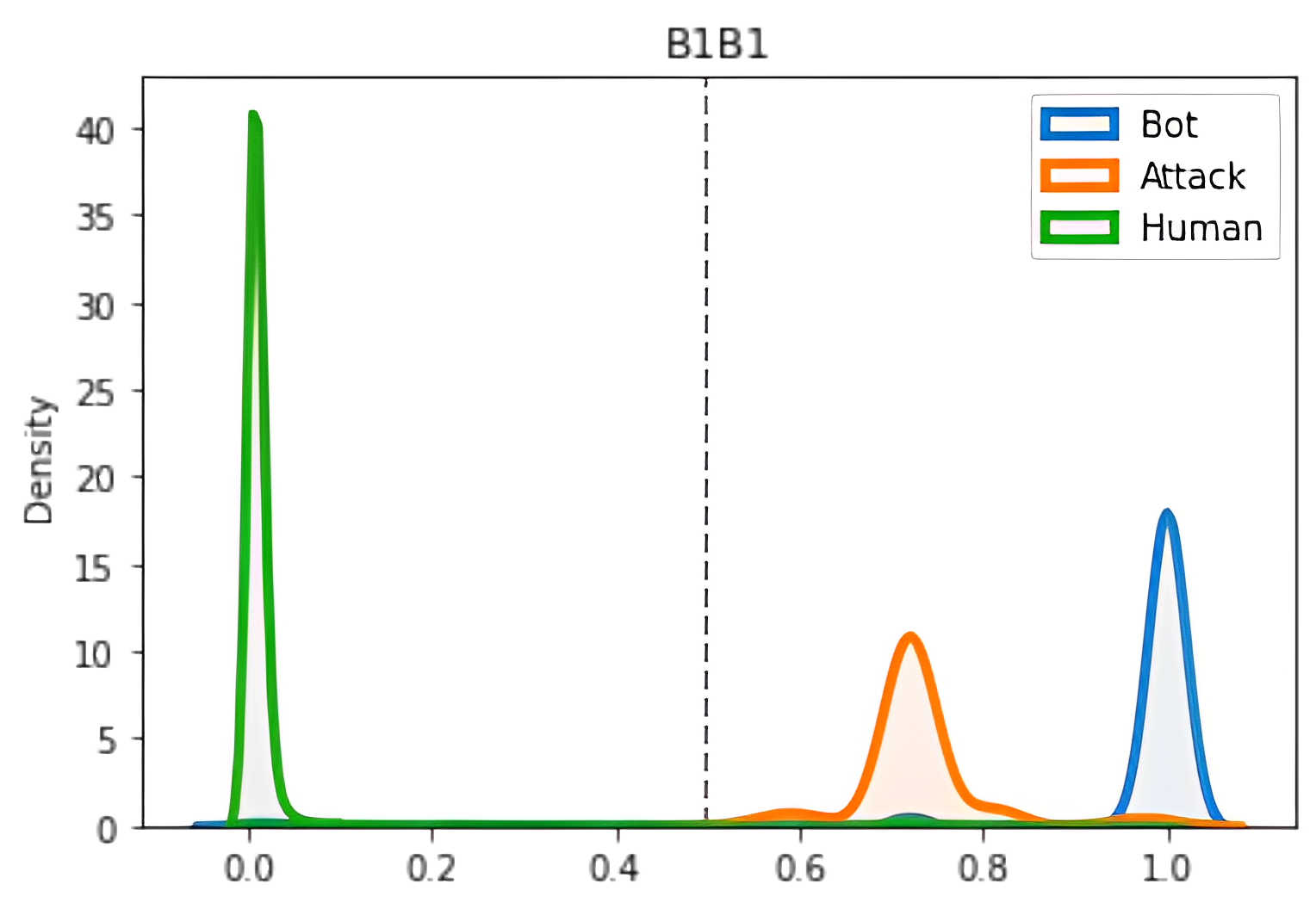}
\caption{ Bot1-Political}
\label{fig:B1last}
\end{subfigure}\hfill
\begin{subfigure}{0.3\textwidth}
    \includegraphics[width=\linewidth]{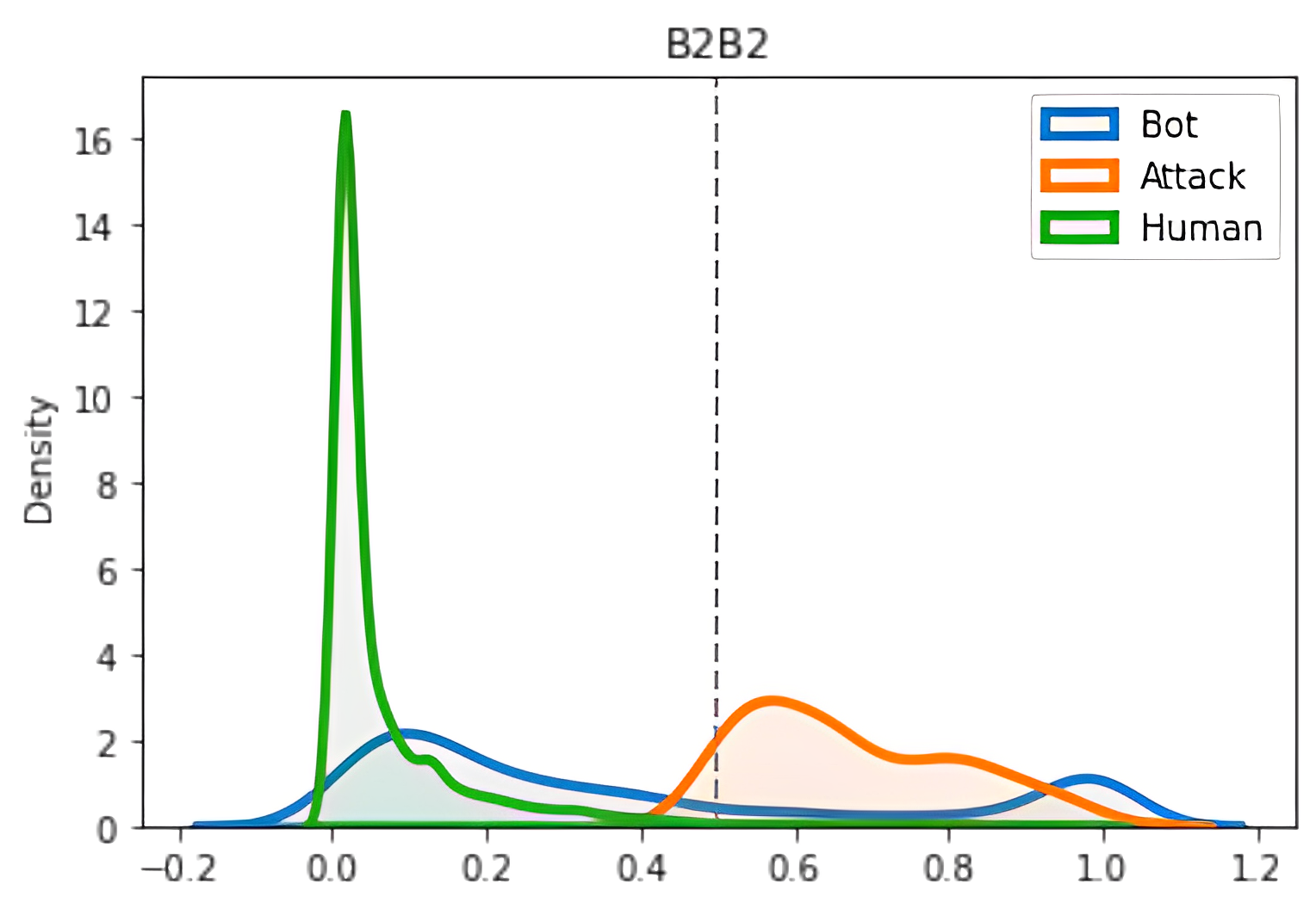}
\caption{ Bot2-Financial}
\label{fig:B2last}
\end{subfigure}\hfill
\begin{subfigure}{0.3\textwidth}
    \includegraphics[width=\linewidth]{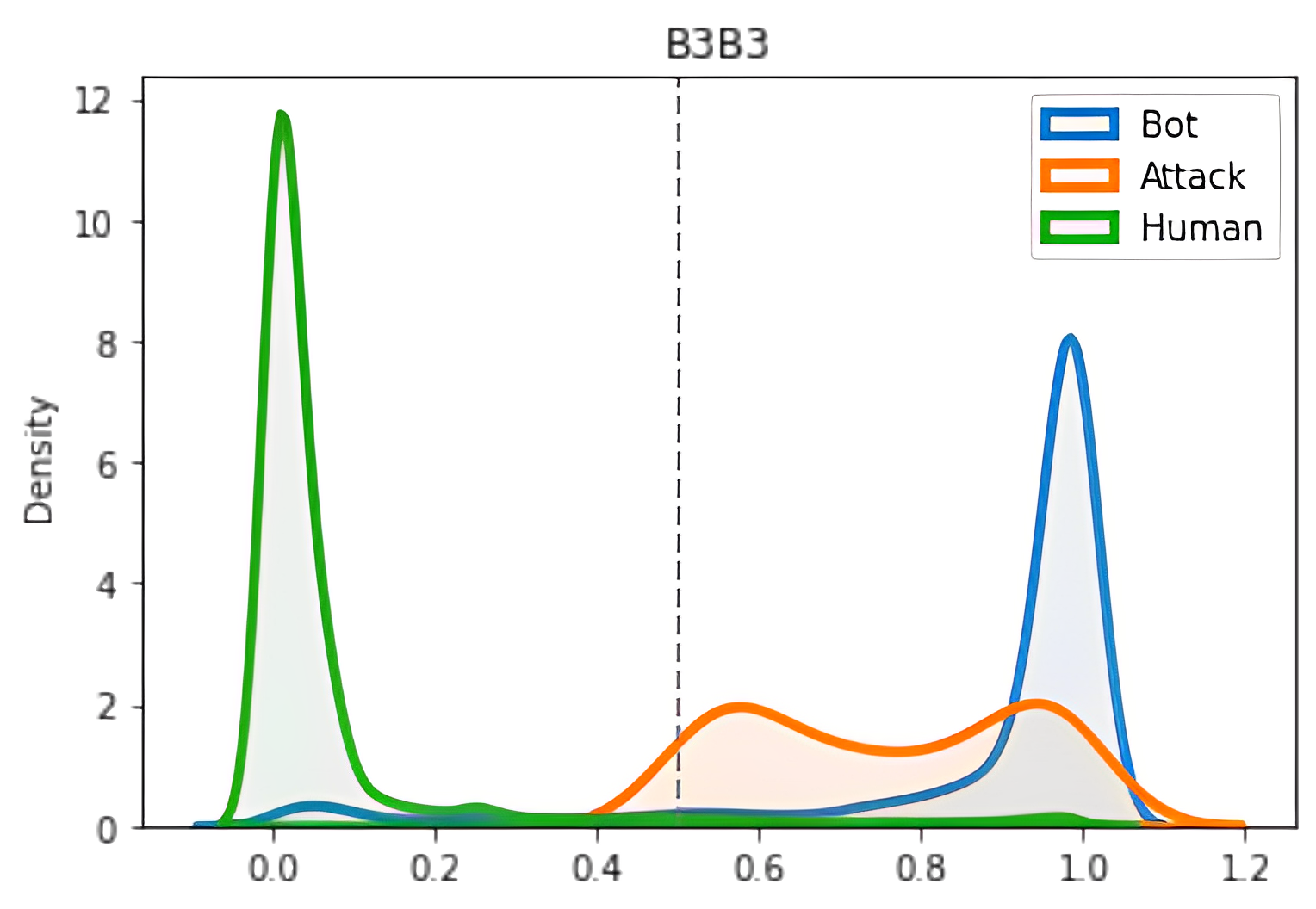}
\caption{ Bot3-Commercial}
\label{fig:B3last}
\end{subfigure}
\caption{Probability distribution of three classes of social bots for attack examples selected from data.}
\label{figure-H-attack}
\end{figure*}
\subsection{scenario 2}

In this part, we evaluate the bot detection model based on the attack examples poisoned to the dataset based on two different scenarios: attack examples are pinching from the human dataset or generating by a social bot.

\subsubsection{Attack examples are pinching from the human's dataset}

In social networks, bots can use deceptive tactics by spreading content written by humans and making other users believe that it comes from a real person. In this evaluation phase, we focus on evaluating the ability of our bot detection model to identify repeat bots that use this strategy on social media platforms.

Bots can use various strategies to mimic human behavior, as mentioned earlier. They can adapt their actions based on the performance of bot detection models and specifically target examples where they are less likely to be detected. In this experiment, we specifically select examples with a low probability of detection to evaluate the effectiveness of the model in detecting these deceptive behaviors.

The results presented in Figure \ref{figure-H-attack} show the probability distribution between three classes: human, bot, and attack examples. The attack examples are derived from the human class and placed in the bot class to simulate deceptive behavior.
As expected, the detection probability differ among the three models trained on different types of bots during the training and testing process. Figure \ref{figure-H-attack} shows that detecting attack examples within the bot class is more difficult for the model trained on commercial bots than for the models trained on political and financial bots. Furthermore, the model trained on financial bots shows greater difficulty in detecting attack examples than the model trained on political bots.

\textbf {Achievement 2:} These results show that attack example detection varies in difficulty across different types of bots. Attack example detection confirm to be more difficult for the model trained on commercial bots, followed by the model trained on financial bots, while the model trained on political bots performs comparatively better.

\begin{figure*}
\begin{subfigure}[t]{0.333\textwidth}
    \includegraphics[width=\linewidth]{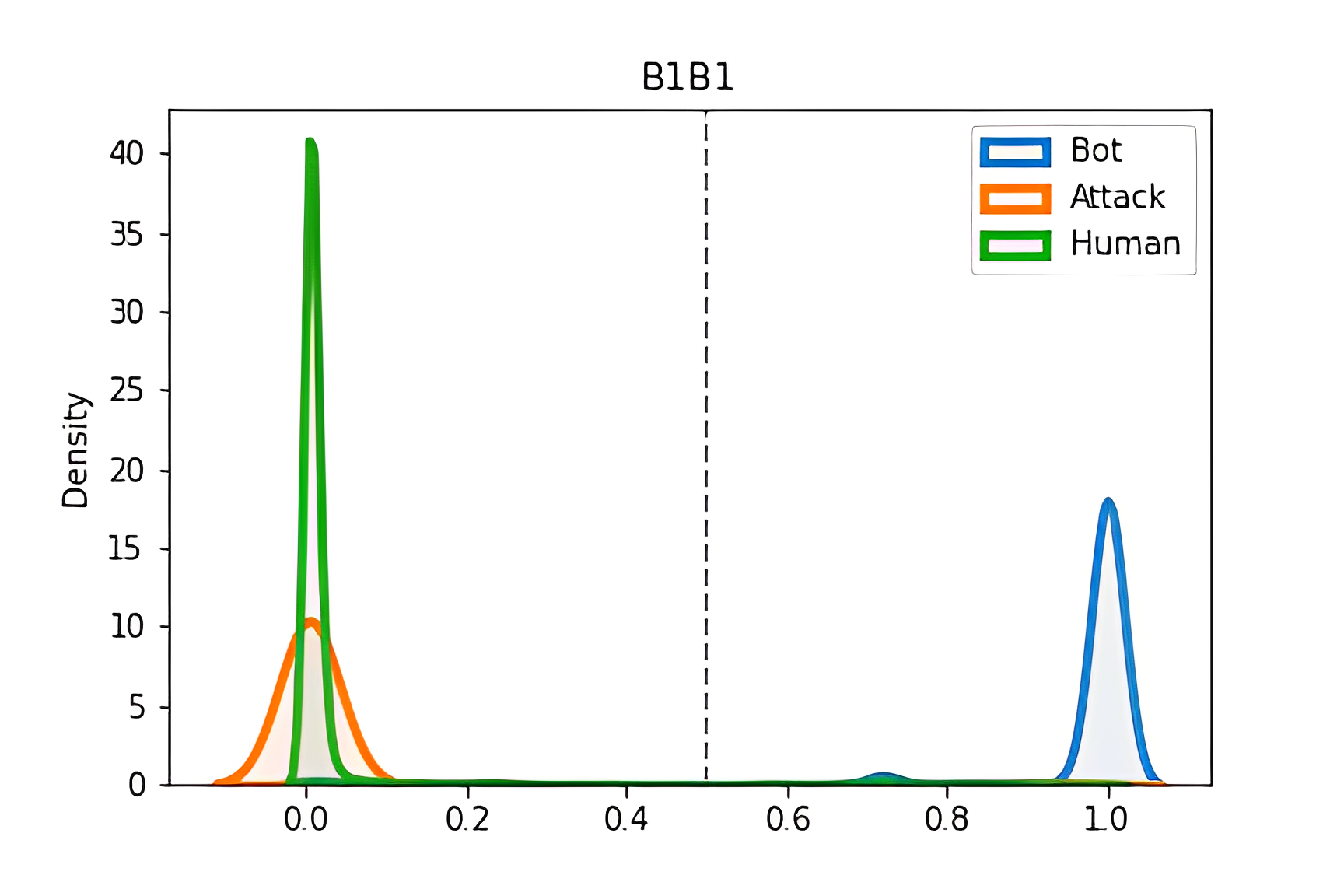}
\caption{Bot1-Political }
\label{fig:Bot1}
\end{subfigure}\hfill
\begin{subfigure}[t]{0.333\textwidth}
    \includegraphics[width=\linewidth]{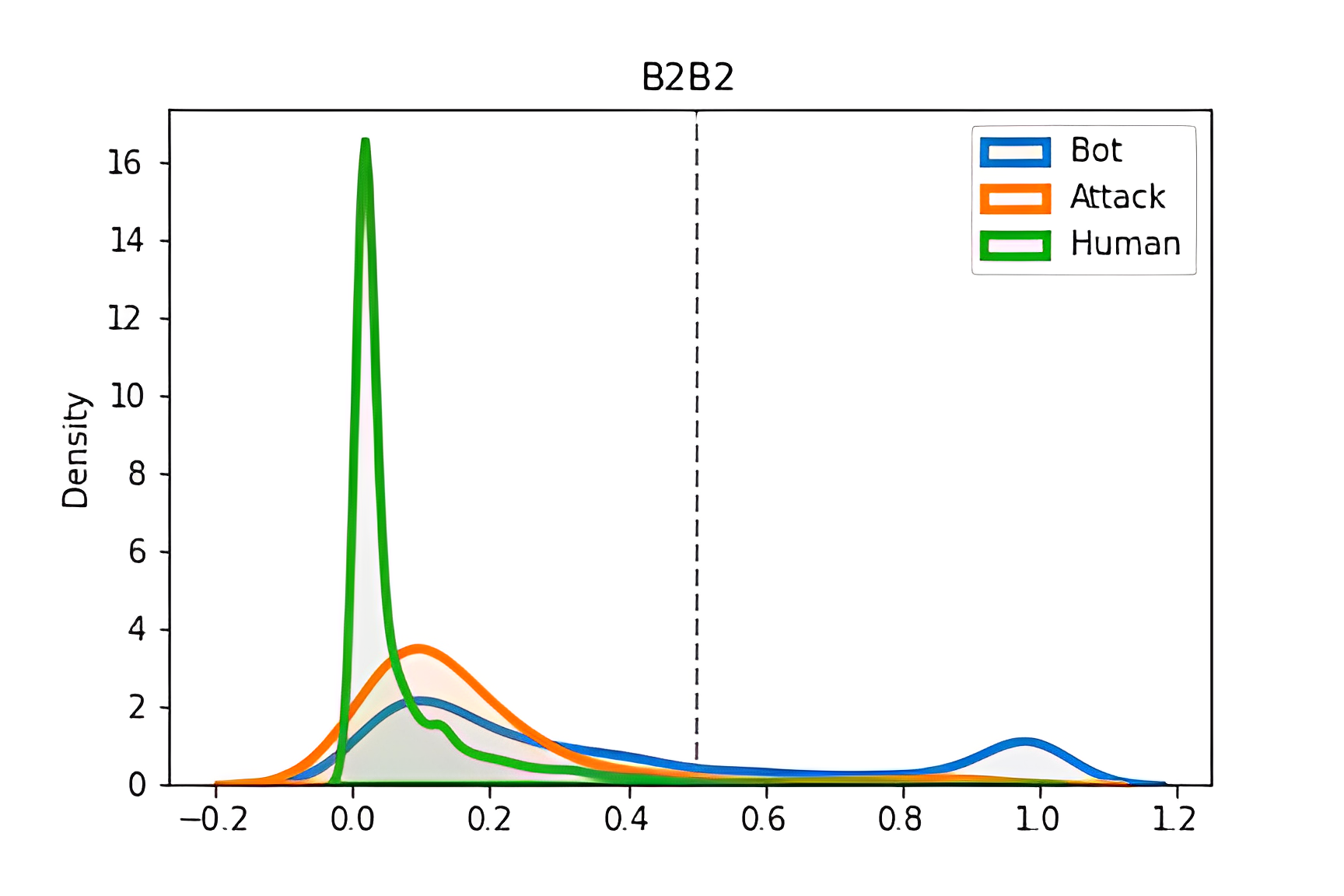}
\caption{Bot2-Financial }
\label{fig:Bot2}
\end{subfigure}\hfill
\begin{subfigure}[t]{0.333\textwidth}
    \includegraphics[width=\textwidth]{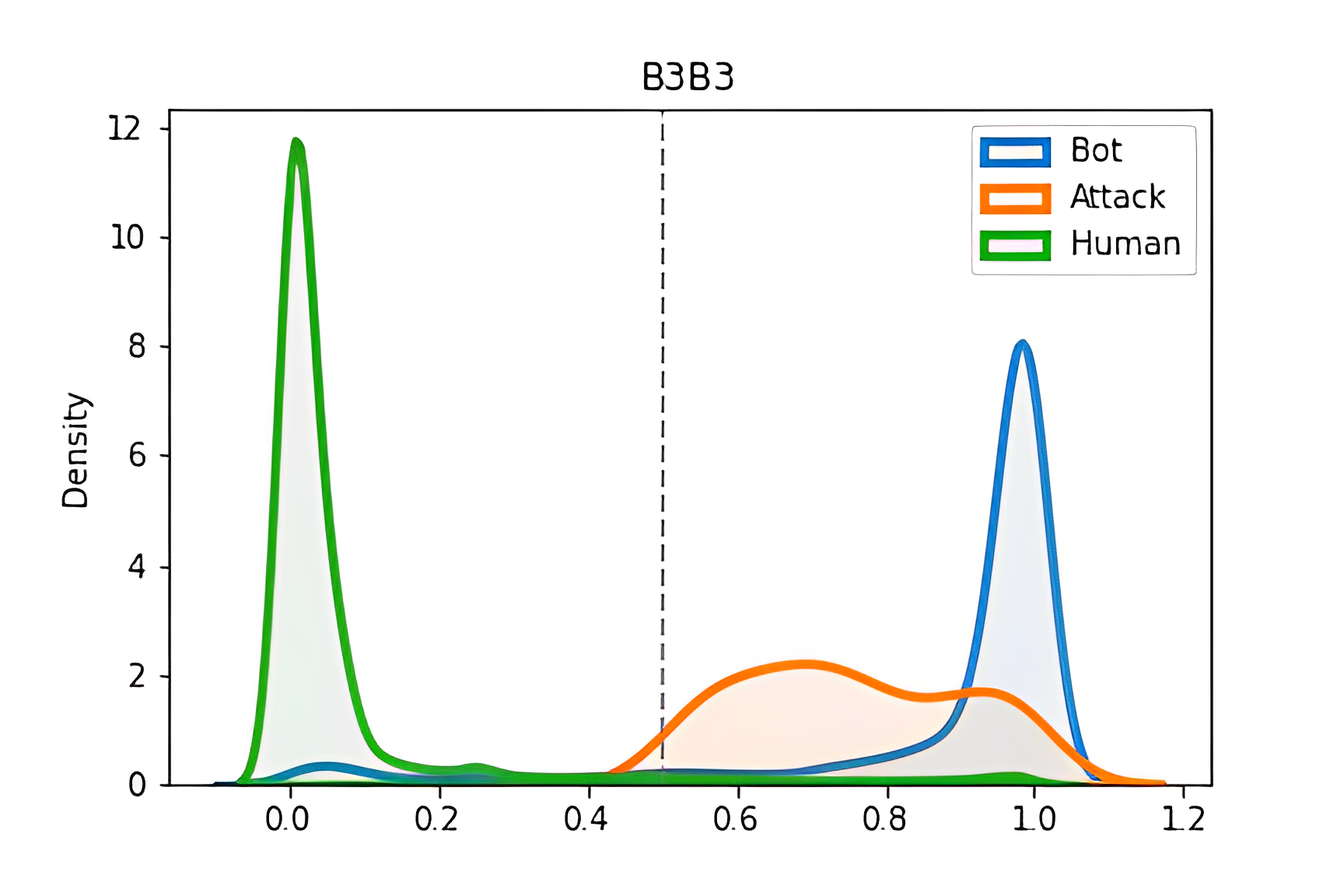}
\caption{Bot3-Commercial }
\label{fig:H-like}
\end{subfigure}
\caption{Probability distribution of three classes of social bots for attack examples generated in adversarial games.}
\end{figure*}
\subsubsection{Attack examples are Generated Human-like samples}
To generate examples that mimic human behavior, we used a generative model trained in an adversarial game, treating it like a bot. These generated examples that resembled human-like interactions were then included in the bot class. A classifier was then used to evaluate the likelihood that these examples belonged to either the bot or human class.

The probability distribution of the bot, human, and attack examples is shown in Figure \ref{fig:H-like}. The results reveal interesting patterns. Models trained on bot examples from the political and financial domains have difficulty distinguishing attack examples from human examples, resulting in predictions that are close to the human class. However, when the generated examples are input to the model trained on commercial bot examples, it exhibits a higher probability of categorizing them as belonging to the bot class. This indicates that the model trained on commercial social bots performs relatively worse than the models trained on political and financial bots.

\textbf {Achievement 3:} These results suggest that the model trained on commercial bots has difficulty in accurately detecting and distinguishing attack samples from human samples, in contrast to the models trained on political and financial bots. The differences in performance suggest differences in the effectiveness of the models in detecting deceptive behaviors and highlight the need for further analysis.
\begin{figure*}
        \begin{subfigure}[ht]{0.32\linewidth}
            \includegraphics[width=4cm, height=5cm]{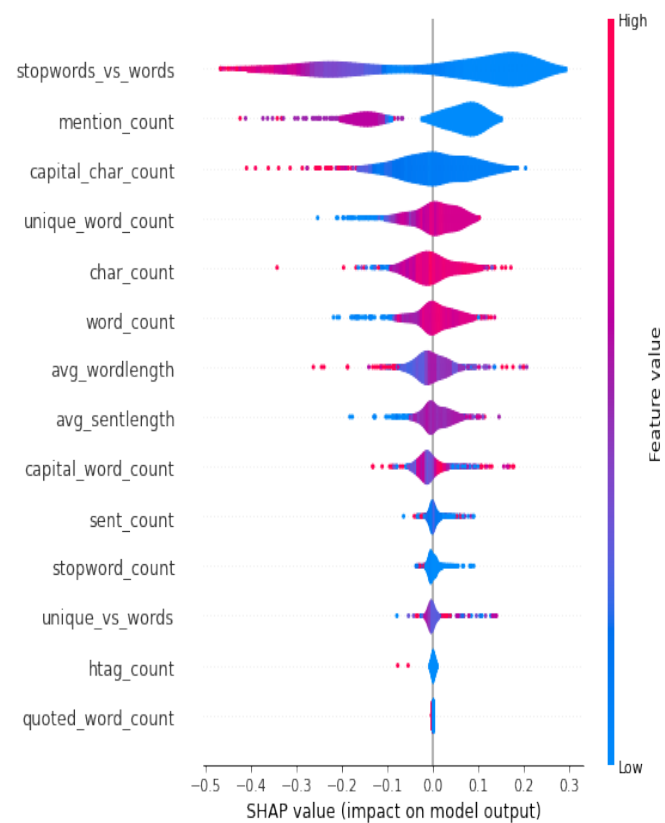}
            \caption{Political bot}
        \end{subfigure}
        \begin{subfigure}[ht]{0.32\linewidth}
            \includegraphics[width=4cm, height=5cm]{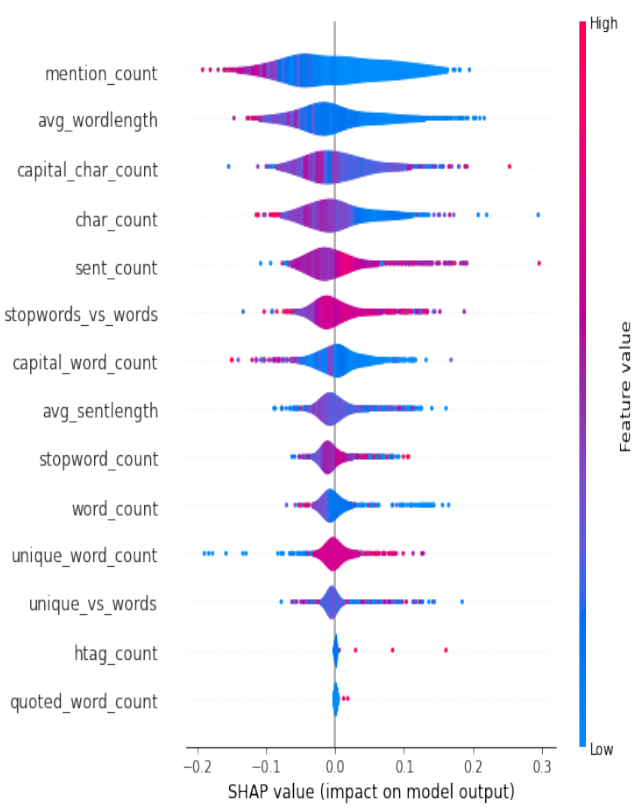}
            \caption{Financial Bot}
        \end{subfigure}
        \begin{subfigure}[ht]{0.32\linewidth}
            \includegraphics[width=4cm, height=5cm]{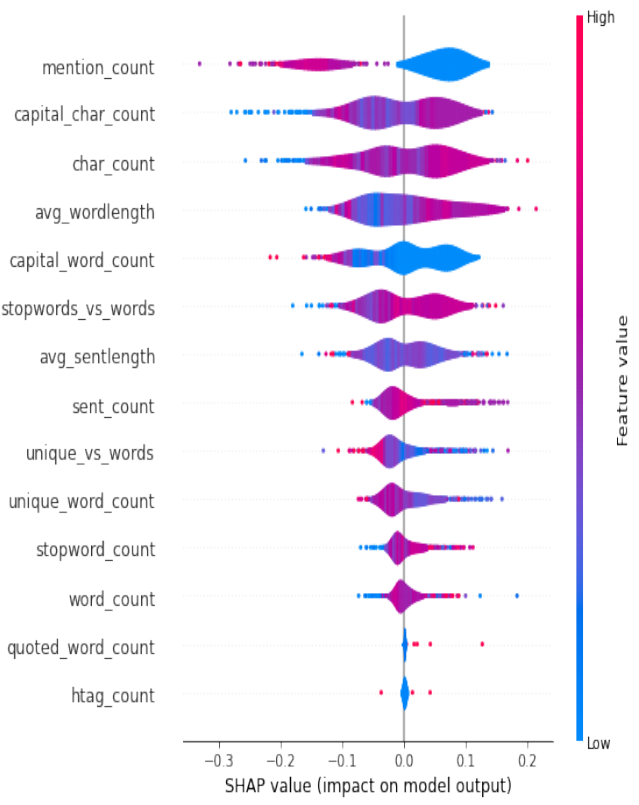}
            \caption{Commercial Bot}
        \end{subfigure}
        \caption{Explain bot detection model based on 14 NLP-based features.}
        \label{fig:Shaps}
\end{figure*}
\subsection{Scenario3}
In this section, we will take a closer look at the textual characteristics of the three different classes of social bots and examine their differentiating features in more detail.

Figure \ref{fig:Shaps} illustrates the relationship between the SHAP values and the prediction performance, with the horizontal axis representing the SHAP value and the color of the points indicating the relative values of the observations compared to those with higher or lower values.

The analysis shows that not all features have the same importance in prediction. For example, the number of mentions has a negative effect on prediction performance when its value is higher. Vice versa, lower mention count values have a positive effect on prediction performance for all three types of social bots.

In the case of political bots, the stopword\_to\_total word ratio feature displays a negative impact when its value is higher, while lower values have a positive impact on the prediction. On the other hand, this feature appears to be less significant for financial and commercial bots compared to other features. Consequently, it may be feasible to exclude this feature during the pre-processing step, given its relatively lower importance.

Furthermore, the results highlight the similarity between the features associated with financial and commercial bots, while political bots show distinct behavioral characteristics in these identified features.

\textbf {Achievement 4:} This analysis provides valuable insights into the importance and influence of different features on the bot detection model's predictions. By understanding the diverse impacts of these features, we can clarify our pre-processing steps and improve the overall accuracy and performance of the bot detection model, particularly when distinguishing between different types of social bots.

%\subsection{Cross-data Analysis}
%We also measured the Pearson correlation between different features of various social bots. In Figure \ref{BB}, we measured the correlation of social bots with each other as well as with the class of human data. On the left side of the figure, financial and commercial bots have the highest correlation, followed by political and financial bots, and finally, political and commercial bots have the lowest correlation. On the right side of Figure \ref{BB}, the correlation between each social bot and human data is shown. Here, the commercial bot has the highest correlation with the human class, while the financial bot has the lowest correlation. Correlation can be seen as a measure of similarity between two datasets  \cite{najari2019link}. When two datasets are similar, it is difficult to separate them,  making it difficult for  a classifier to distinguish between the two classes. As a result, it is likely that predicting attacks using  a model trained on a commercial bot data will be more challenging than using a model trained on financial or political bot data.
%\begin{figure}
%        \begin{subfigure}[ht]{0.7\linewidth}
%            \includegraphics[width=8cm, height=5cm]{BB.png}
%            \caption{Bot-Bot}
%            \label{BB}
%        \end{subfigure}
%        \begin{subfigure}[ht]{0.7\linewidth}
%            \includegraphics[width=8cm, height=5cm]{HB.png}
%            \caption{Human-Bot}
%            \label{HB}
%        \end{subfigure}
%        \caption{Pearson Correlation of Bot-Bot and Human-Bot.}
%        \label{fig:correlation}
%\end{figure}
There have been some studies that aim to conduct a cross-domain analysis, in which the goal is to evaluate the generalization of a model by training it on a dataset from one domain and testing it on a dataset from a different domain\cite{yang2020scalable}. 
\begin{figure}
	\begin{center}
		\includegraphics[width=0.6\linewidth]{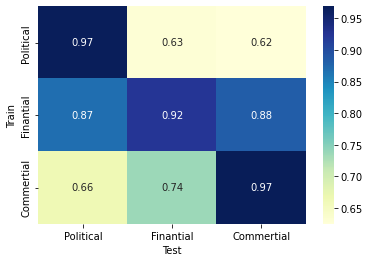}
		\caption{Accuracy values for cross-data analysis.}
		\label{fig:Corresponding}
	\end{center}
\end{figure}
To further assess the performance of a text-based classifier, we performed an evaluation where the classifier was trained on the human class and a specific social bot, and then tested on a human test set and another social bot. The outcomes of this evaluation are shown in Fig. \ref{fig:Corresponding}.

The results demonstrate that the model exhibited poor performance when predicting the political social bot using a model trained on the commercial bot. This lack of accuracy can be attributed to the lowest correlation between the two domains. On the contrary, the model performed relatively similarly when predicting the commercial and financial bots using a model trained on the political bot, or when predicting the political bot using a model trained on the political bot itself.

This results indicate that the prediction of political bots using models trained on commercial and financial bots yields a higher accuracy. However, when the model trained on the financial bot is tested on the same dataset, its accuracy is lower compared to the other bot datasets. This discrepancy may be attributed to the fact that the model trained on the financial dataset exhibits better generalization capabilities, while the models trained on the commercial and political datasets are more prone to over-fitting on the training set, resulting in worse performance in different circumstances.

Also, these findings suggest that the classifier may be more effective at predicting political bots using models trained on commercial and financial bots. However, it may not generalize well to other scenarios or domains.

\textbf {Achievement 5:} Here, our findings indicate that the classifier may not be highly effective in accurately predicting social bots from different domains. The divergent characteristics and behaviors exhibited by social bots in different domains pose challenges for the classifier in effectively capturing and distinguishing their distinguishing features.

\subsection{Comparison}
\begin{figure}
	\begin{subfigure}[ht]{0.5\linewidth}
		\includegraphics[width=4.5cm, height=4cm]{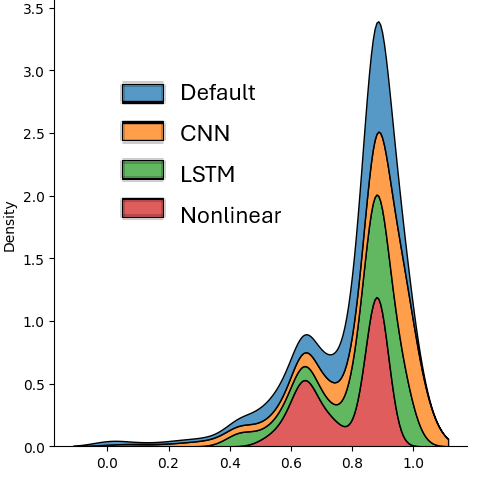}
		\caption{Probability distribution for class bot}
	\end{subfigure}
	\begin{subfigure}[ht]{0.5\linewidth}
		\includegraphics[width=4.5cm, height=4cm]{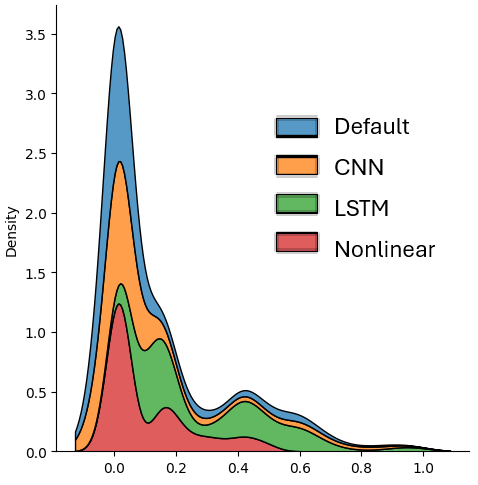}
		\caption{Probability distribution for class human}
	\end{subfigure}
	\caption{ Model Comparison - The Probability distribution of two class bot and human in different models.}
	%  \reza{add x y axes}
	\label{fig:model-compare}
 %\reza{legend is not easy to read, pls increase the font and move them to top, for (a) top left pls}
\end{figure}
In this section, we have evaluated the effectiveness of four different approaches. %To accomplish this, we employed the BERT embedding method to convert the input text into vectors.

Since the generation of word embeddings from Bidirectional Encoder Representations from Transformers (BERT), a pre-trained language model, is an efficient technique, we used this method together with neural networks. Therefore, we used four models for comparison: CNN, LSTM, Nonlinear and Default. The default model relies exclusively on BERT for predictions.

Fig \ref{fig:model-compare} shows the probability distribution of two classes, bot and human, for the Cresci dataset. It is noticeable that the probabilities for the bot class tend towards 1, while the probabilities for the human class tend towards 0. The default model has the highest distribution for both classes, followed by the CNN, LSTM and Nonlinear models in successive order. Table \ref{tb:models} shows the performance of these four models in two datasets, Cresci-2017 and Midterm-2018, confirming the previous results.

\begin{table}
	\begin{center}
		\caption{Dataset Comparison - Cresci 2017 \cite{cresci2017paradigm} Vs. Midtrerm 2018 \cite{yang2020scalable} (Cresci/Midterm).}
		%\reza{cite Cresci in the caption pls}
		\label{tb:models}
		%\resizebox{\columnwidth}{!}{%
			\begin{tabular}{c|c|c}

				\textbf{Model} &\textbf{Accuracy} & \textbf{AUCROC}\\
				\hline
				Default  & 0.915/ 0.825 & 0.971/ 0.588 \\
				CNN &0.893/ 0.799  & 0.962/ 0.704\\
				LSTM  & 0.819/ 0.825  & 0.902/ 0.522 \\
				NonLinear  & 0.875/ 0.825 & 0.951/ 0.523 \\
			\end{tabular}
		%}
	\end{center}
\end{table}

%\section{Achievements for Future Works}
%When we evaluate a model trained on a dataset, the opponent from the test set, we can see the model trained on a financial dataset has reached higher accuracy. Its reason is that the model trained on the financial dataset has a higher generalization than the others, while the models trained on commercial and political datasets are over-fitted on the train set and cannot act in the other circumstances so well.

	\section{Conclusions}\label{sec:Conclusions}
This research employed a synthetic adversarial game to assess the effectiveness of text-based social bot detection methods in different scenarios. The focus was on evaluating the performance of generative models in generating attack examples that mimic human textual behavior. The findings revealed that these detection methods exhibit varying levels of strength across different types of social bots, with performance variations based on the specific domain and content generated by the bots. %Particularly, political bots should not be considered as reliable references for predicting other types of bots, as their textual behavior significantly differs from that of commercial and financial bots.

In future works, the aim is to conduct more extensive  bot detection models based on this works achievements. By gaining deeper insights into the behavioral patterns of social bots, the research aims to enhance the performance of detection models. This will involve refining the understanding of bot behavior and exploring novel approaches to improve detection accuracy and effectiveness.

By addressing the limitations identified in this study and further investigating the complexities of social bot detection, the research endeavors to advance the field and contribute to the development of more robust and reliable methods for detecting and combating social bots.

	%\section{Acknowledgments}\label{sec:Acknowledgments}
%Mostafa Salehi is supported by a grant from IPM, Iran (No. CS1402-4-268).
%\reza{I suggest we remove this ack for submission, and add it in the final version when it is accepted (most of the venues are a bit sensitives these days ...).}

  \bibliographystyle{IEEEtran}
  \bibliography{8_citation}

		%\bibliography{bibliography}
\end{document}